\documentclass[12pt,a4paper]{article}

\usepackage{hyperref}
\usepackage[english]{babel} 
\usepackage{amsthm}
\usepackage{amsmath}
\usepackage{amsfonts}
\usepackage{amssymb}
\usepackage{enumerate}
\usepackage{amscd}
\usepackage{graphicx}
\usepackage{makeidx}
\usepackage[all]{xy}

\makeindex

\newcounter{Def}[section]

\theoremstyle{plain}

\newtheorem{Example}[Def]{Example}

\newenvironment{Definition}{ \refstepcounter{Def}\mbox{}\\ 
\noindent\sl\textbf{Definition
\arabic{section}.\arabic{Def}} }
{\vspace{0.3cm}}

\usepackage[all]{xy}

\xyoption{2cell}
\makeindex
\begin{document}

\numberwithin{equation}{section}

                         %      \version\versionno  \draft

\thispagestyle{empty}
\begin{flushright}
   %{\sf ZMP-HH/10-12}\\
   {\sf Hamburger$\;$Beitr\"age$\;$zur$\;$Mathematik$\;$Nr.$\;$425}\\[2mm]
   November 2011
\end{flushright}
\vskip 2.0em
\begin{center}\Large
Bicategories in field theories - an invitation
\end{center}\vskip 1.4em
\begin{center}
  Thomas Nikolaus$^1$ and Christoph Schweigert$^2$
\end{center}

\vskip 3mm

\begin{center}\it
$^1$ Fakult\"at f\"ur Mathematik,
Universit\"at Regensburg \\
Universit\"atsstra\ss{}e 31,
93053 Regensburg \\[.3em]
$^2$  Fachbereich Mathematik, \ Universit\"at Hamburg\\
  Bereich Algebra und Zahlentheorie\\
  Bundesstra\ss e 55, \ D\,--\,20\,146\, Hamburg
\end{center}
\vskip 2.5em
\begin{abstract} \noindent
We explain some applications of bicategories in both classical and quantum
field theory. This includes a modern perspective
on some pioneering work of Max Kreuzer and Bert Schellekens on rational 
conformal field theory.

\end{abstract}

\section{Introduction}\label{tncs_sec1}

The notion of a category undeservedly has the reputation of being
abstract. In fact, it just summarizes the combinatorics of arrows.
Ignoring set theoretic
subtleties, a category consists of two entities: first, objects, which can be represented by
points. Second, morphisms are then represented by arrows starting or ending at these points.
The basic operation is concatenation of arrows.
This composition is required to be associative and
to have neutral elements. Examples of categories are abundant; the following examples are
of particular importance for us:

\begin{Example}\mbox{} 
\begin{enumerate}
\item
The category $\mathrm{Vect}_k$ of vector spaces over a given field $k$: 
objects are vector spaces,  morphisms are linear maps.
\item
The category of, say complex, 
algebras: objects are complex algebras, morphisms are
morphisms of algebras. 
\item The category of smooth oriented cobordisms $\mathrm{Cobord}_{d+1,d}$: objects are 
smooth oriented $d$-dimensional manifolds without
boundaries. A morphism $\Sigma_1\to\Sigma_2$ is a (diffeomorphism class of an) oriented
$d+1$-dimensional manifold $M$, together with a diffeomorphism 
$\Sigma_1\sqcup \overline{\Sigma_2}\to\partial M$.
\end{enumerate}
\end{Example}

Since categories have %two layers of structure, 
objects and morphisms, functors
are defined as pairs of compatible maps, on objects and on morphisms. The notion of a
functor allows to give a concise definition of a $(d+1,d)$-dimensional {\em topological field theory}
that is originally due to Atiyah \cite{At}: it is a (symmetric monoidal) functor
$\mathrm{Cobord}_{d+1,d}\to\mathrm{Vect}$.

Let us now introduce the central notion of this contribution, the notion of a bicategory.
For categories, the underlying combinatorics is given by dots and arrows; for
bicategories, the combinatorics of {\em bigones} is relevant:

$$\xy
(-8,0)*+{a\quad\bullet\quad}="4";
(8,0)*+{\quad\bullet\quad b}="6";
{\ar@/^1.65pc/^{f} "4";"6"};
{\ar@/_1.65pc/_{g} "4";"6"};
{\ar@{=>}^<<<{\scriptstyle \varphi} (0,3)*{};(0,-3)*{}} ;
\endxy $$

We have now three layers of structure: objects $a,b,\ldots$, 1-morphisms
$f,g\ldots$, and 2-morphisms $\varphi$.
1-morphisms can be concatenated, in the same way as morphisms in a category:
$$ \xymatrix{
a\quad \bullet \ar^f[r]& \bullet\quad b \ar^g[r]&\bullet \quad c
}\quad\mapsto\quad 
\xymatrix{
a\quad \bullet \ar^{f\otimes g}[r]& \bullet\quad c}
$$
We denote the corresponding concatenation with $\otimes$;
this
concatenation is not necessarily strictly associative - an issue we will
not discuss in this contribution. 
For 2-morphisms, there are two different concatenations: vertically
$$
\xy
(-8,0)*+{a \quad\bullet\quad}="4";
(8,0)*+{\quad\bullet\quad b}="6";
{\ar "4";"6"};
{\ar@/^1.75pc/^{f} "4";"6"};
{\ar@/_1.75pc/_{g} "4";"6"};
{\ar@{=>}^<<{\scriptstyle \alpha} (0,6)*{};(0,1)*{}} ;
{\ar@{=>}^<<{\scriptstyle \beta} (0,-1)*{};(0,-6)*{}} ;
\endxy
\quad\mapsto\quad
\xy
(-8,0)*+{a\quad\bullet\quad}="4";
(8,0)*+{\quad\bullet\quad b}="6";
{\ar@/^1.65pc/^{f} "4";"6"};
{\ar@/_1.65pc/_{g} "4";"6"};
{\ar@{=>}^<<<{\scriptstyle \alpha\cdot\beta} (0,3)*{};(0,-3)*{}} ;
\endxy 
$$
and horizontally:
$$
\xy
(-16,0)*+{\bullet}="4";
(0,0)*+{\bullet}="6";
{\ar@/^1.65pc/^{f} "4";"6"};
{\ar@/_1.65pc/_{g} "4";"6"};
{\ar@{=>}^<<<{\scriptstyle \alpha} (-8,3)*{};(-8,-3)*{}} ;
(0,0)*+{\bullet}="4";
(16,0)*+{\bullet}="6";
{\ar@/^1.65pc/^{f'} "4";"6"};
{\ar@/_1.65pc/_{g'} "4";"6"};
{\ar@{=>}^<<<{\scriptstyle \beta} (8,3)*{};(8,-3)*{}} ;
\endxy
\quad\mapsto\quad
\xy
(-8,0)*+{\bullet}="4";
(8,0)*+{\bullet}="6";
{\ar@/^1.65pc/^{f\otimes f'} "4";"6"};
{\ar@/_1.65pc/_{g\otimes g'} "4";"6"};
{\ar@{=>}^<<<{\scriptstyle \alpha\otimes\beta} (0,3)*{};(0,-3)*{}} ;
\endxy 
$$
Since bicategories have three layers of structures, it is clear that bi-functors
relating two bicategories should act on objects, 1-morphisms and 2-morphisms.

We give examples of bicategories that will be relevant for us:
\begin{Example}\mbox{}
\label{ex1}
\begin{enumerate}
\item
The bicategory $2\mathrm{Vect}$ of 2-vector spaces: objects are $\mathbb C$-linear finitely
semi-simple abelian categories. 
%(Bimodules over algebras are in fact a full subcategoryof this category.)
%tn: Ich denke, das stimmt nicht. Eher ist diese Kategorie eine volle Unterkategorie von Algebren, denn warum sollte jede Algebra halbeinfach sein!? Andersrum können wir immer die Algebra C^n nehmen.
%cs: Sie haben natürlich recht. Vorschlag: Klammer streichen, Reihenfolge der Beispiele umdrehen. Habe
%     Ich schon gemacht.
% TN: Ich hab jetzt noch einen Vorschlag gemacht unten.
\item
The bicategory of complex algebras with algebras as objects, bimodules
as 1-morphisms and intertwiners of bimodules as 2-morphisms. The
concatenation of 1-morphisms is the tensor product of bimodules over the appropriate
algebra, $B\otimes_A B'$.
(The bicategory $2\mathrm{Vect}$ from the first example is in fact equivalent to the full subcategory of algebras isomorphic to $\mathbb{C}^n$.)

\item \label{ex1_third}
Using ideas from 2-form gauge theories, we assign to any smooth manifold $M$
a bicategory $Grbtriv^\nabla(M)$. Its objects are 2-forms $B\in\Omega^2(M)$,
i.e.\ gauge fields, its 1-morphisms $B\stackrel\lambda\to B'$ are 1-forms
$\lambda$ such that $\mathrm{d}\lambda=B'-B$, i.e.\ gauge transformations parametrized by
1-forms.
The 2-morphisms are gauge transformations of gauge transformations:
$\mathrm{U}(1)$-valued functions $\varphi$ on $M$
such that $\frac1{\mathrm{i}}\mathrm{d}\log\varphi=\lambda'-\lambda$.

\item
Our final example are extended cobordisms $\mathrm{Cobord}_{3,2,1}$.
Its objects are oriented smooth one-dimensional manifolds, i.e.\ disjoint unions
of circles. 1-morphisms are cobordisms of these one-dimensional manifolds.
For example  the pair of pants,
i.e. the three-punctured sphere, can be seen as a 1-morphism from two circles to a
single circle. 2-morphism finally are manifolds with corners. 
Figure 1 shows an example of a 2-morphism.
\begin{figure}[h!]
\centerline{\scalebox{1.0}{\includegraphics{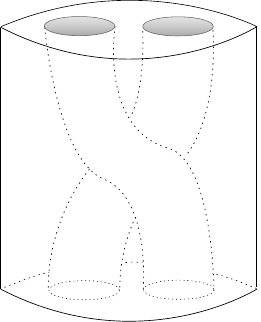}}}
\caption{A 2-morphism between two pair of pants, i.e.\ two three-punctured spheres.}
\label{dose_fig}
\end{figure}
\end{enumerate}
\end{Example}
\section{Equivariant Hopf algebras from topological field theories}\label{tncs_sec2}

We start by describing  an application of bicategories to topological
field theory, yielding interesting algebraic and representation theoretic structures.
The examples (2) and (4) of bicategories  suggests the following

\begin{Definition}
An \emph{extended topological field theory} is a (symmetric monoidal) bifunctor
$$ Z: \mathrm{Cobord}_{3,2,1} \to 2\mathrm{Vect} \,\, . $$
\end{Definition}

The qualifier ``extended'' can be justified by the observation that
$Z$ can be restricted to an ordinary functor from the endomorphism category of the object
$\emptyset$ to the category of vector spaces. Unpacking the definitions, one finds
that this restriction is a topological field theory in the sense of Atiyah.

Dijkgraaf-Witten theory \cite{dw90} is an example of an extended topological field theory. The following
explicit construction is based on the contributions of many authors, in particular
of Freed-Quinn \cite{FQ93} and of Morton \cite{Mor10}. Let $G$ be a finite group;
for a smooth manifold $M$ of any dimension, denote by $\mathcal{A}_G(M)$ the category of $G$-bundles on $M$
which is actually a groupoid, i.e.\ a category in which all morphisms, namely all
gauge transformations, are invertible.
From this structure, on can build a bifunctor $\tilde{\mathcal A}_G$ from $ \mathrm{Cobord}_{3,2,1}$ to
the category of spans of groupoids. Dijkgraaf-Witten theory is then the concatenation
of $\tilde{\mathcal A}_G$ with a 2-linearization:
$$ Z_{DW}: \mathrm{Cobord}_{3,2,1}\stackrel{\tilde{\mathcal A}_G}\to 
\mathrm{SpanGrp} \stackrel{[-,\mathrm{Vect]}}\to 2{-}\mathrm{Vect} \,\, . $$

Let us evaluate this functor explicitly in examples:
\begin{itemize}
\item The 1-manifold $S^1$ is sent to the category $\mathcal{C}:=[\mathcal{A}_G(S^1),\mathrm{Vect}]$.
Taking into account that the category $\mathcal{A}_G(S^1)$ is equivalent to the action groupoid
$G//G$ for the conjugation action (this is of course just ``holonomy modulo
gauge transformations at a point), one easily deduces that
this category is equivalent to the category of representations of the
Drinfeld double, $\mathcal{C}\cong\mathcal{D}(G){-}\mathrm{mod}$.

\item
The pair of pants, i.e.\ the sphere with two ingoing and one outgoing puncture
yields a functor $\mathcal{C}\boxtimes\mathcal{C}
\to\mathcal{C}$ which gives rise to a tensor product.

\item
2-morphisms, i.e.\ 3-manifolds with corners, finally give rise to
natural transformations. For example, the manifold with corners depicted in 
figure \ref{dose_fig} gives rise to a braiding on the category $\mathcal C$.
\end{itemize}

This construction can be extended \cite{jmtncs} to an equivariant version:
Let $J$ be another finite group (that is not a priori related to the group $G$). 
A weak $J$-action on the stack ${\mathcal A}_G$ amounts
to a weak $J$-action on the group $G$, i.e.\ a group automorphism $\varphi_j:G\to G$
for any $j\in J$ and an element $c_{j,k}\in G$ for any pair $j,k\in J$ such that
$\varphi_j\circ\varphi_k=\mathrm{ad}_{c_{jk}} \circ \varphi_{jk}$, i.e.\
such that the group law of $J$ is realized only up to inner automorphisms of $G$.
(The elements $c_{jk}\in G$ have to obey appropriate coherence conditions which we
omit here.)

By Schreier's \cite{schreier} results, this amounts to an extension of groups
$$ 1\to G\to H \to J \stackrel\pi\to 1 $$
that is not necessarily split. (One should also keep in mind that the group $G$ is 
not required to be abelian.)

One can now generalize \cite{jmtncs} the construction just described to obtain a $J$-equivariant
extended topological field theory: to any $J$-cover $P\stackrel J\to M$, one associates a category 
${\mathcal A}_G(P\stackrel J\to M)$ of $G$-bundles twisted by $P$. A $G$-bundle
twisted by the cover $P\stackrel J\to M$ is a pair $(Q,\varphi)$, consisting of an 
$H$-bundle $Q$ over $M$ and a smooth map $\varphi: Q \to P$ over $M$ that is required to obey
$\varphi(q · h) = \varphi(q) \cdot \pi(h)$ for all $q\in Q$ and $h\in H$. 
Put differently, a $P \stackrel J\to M$-twisted G-bundle is a lift of the
$J$-cover $P$ along the group homomorphism $\pi : H \to J$. The next steps in the 
construction are, exactly as in \cite{Mor10}, the extension to spans of groupoids and
2-linearization. In this way, one obtains a $J$-equivariant topological field theory from which one
can extract in particular a (weakly) $J$-modular Hopf algebra. 

It should be noted that the normal subgroup $G$ of $H$ gives rise to a crossed module.
In fact, it can be expected that more general crossed modules give rise to
equivariant topological field theories as well, but details remain to be worked out.

\section{2-stacks and gerbes}\label{tncs_sec3}

\subsection{2-stacks}

While the previous application was about the construction of specific bicategories, we
now turn to sheaves of bicategories. Our goal is to associate to every smooth
manifold $M$ a bicategory in such a way that for each diffeomorphism,
there is a pullback functor for the bicategories  
that generalizes the way bundles can be pulled back. Moreover,
we impose gluing conditions for open covers (and generalizations thereof).
Our starting point is Example \ref{ex1} \eqref{ex1_third}. This example has pullbacks, but still does not have all properties we wish to have:
the problem with the bicategory $Grbtriv^\nabla(M)$ can be understood
by considering the related category $Buntriv^\nabla(M)$ of trivial $\mathrm{U}(1)$-bundles
with connection on $M$. It has global 1-forms $A\in\Omega^1(M)$ as objects and
gauge transformations $\frac1{\mathrm i}{\mathrm d} \log \varphi= A'-A$ as
morphisms $A\stackrel\varphi\to A'$. Bundles are only {\em locally} trivial, and
can be glued together to globally non-trivial bundles. Global properties of bundles with
connections have numerous well-known physical implications; global properties are
equally important for the objects we are about to construct.

In the case of 2-form gauge theories, the correct global geometric objects are also
obtained by gluing together  locally trivial objects:
\begin{itemize}
\item The first datum is an open cover $(U_i)$ of the manifold $M$.
\item We then chose a trivial object $B_i\in Grbtriv^\nabla(U_i)$ for each open subset.
\item The third datum is a 1-morphism $A_{ij}: B_j\to B_i$ in $Grbtriv^\nabla(U_i\cap U_j)$
for each intersection $U_i\cap U_j$.
\item The next datum is a choice of 2-morphisms on threefold intersections which are
required to obey coherence conditions on fourfold intersections. For details, we refer to
\cite{tncs}.
\end{itemize}
In this way, we obtain an object called an ($\mathrm{U}(1)$-)bundle gerbe with connection.

The whole construction can be described more systematically as follows: we start with
a bifunctor $Grbtriv^\nabla$ from the (opposed) category
of smooth manifolds to bicategories. It is compatible with pullbacks and thus forms a 2-prestack.
Recall from geometry that sheafification allows to construct to any presheaf
$\mathcal X$ a sheaf ${\mathcal X}^+$.
One of our results, Theorem 3.3 of \cite{tncs} asserts that this plus construction 
can be extended to 2-prestacks. In fact, applying the plus construction to the 
2-prestack $Grbtriv^\nabla$ yields just bicategories of bundle gerbes with connection,
$$ (Grbtriv^\nabla M)^+ = Grb(M) \,\, , $$
as defined in \cite{stevenson,waldorf}.
Theorem 3.3 of \cite{tncs} thus shows that the latter form a sheaf on the
category of smooth manifolds (with the topology given by surjective submersions).

This allows us to construct gerbes by gluing local data, even along surjective
submersions. We have thus the familiar tools from theory of bundles
with connection at our disposal.

The plus construction for 2-stacks relies on quite a few more results. Along the way,
we get results of independent interest, both in mathematics and mathematical physics.
In particular, our results apply not only to manifolds, but 
to Lie groupoids, i.e.\ orbifolds, as well. Since this includes the case of action groupoids,
we also obtain equivariant versions of all our results. As a central result, we mention
Theorem 2.16 of \cite{tncs}: the category ${\mathcal X}(\Lambda)$ associated by a 2-stack 
$\mathcal X$ to a Lie groupoid $\Lambda$ is invariant under Morita equivalence of Lie groupoids.

\subsection{Jandl gerbes}

A bundle with connection on a smooth manifold $M$ leads to the notion of holonomy which associates a group
element to any closed curve in $M$. The action of electrically charged particles
moving in an electromagnetic background field described by the bundle contains terms that can be
formulated in terms of the holonomy.
Bundle gerbes with connection similarly lead to the notion of a
surface holonomy: given an abelian bundle gerbe on $M$, one can associate
to a closed oriented surface in $M$ an element of the group $\mathrm{U}(1)$. Surface holonomies are
in fact one motivation to study bundle gerbes. They also have a physical application:
they are (exponentials of) Wess-Zumino terms and enter in the action of strings moving in
the background described by the bundle gerbe, i.e.\ in a non-trivial $B$-field.

It is by now well-established that sigma models, for example on compact Lie groups, can also
be considered for unoriented surfaces. In fact, they play an important role in type I string
compactifications. This raises the following question: what geometric data on $M$ are needed
to define unoriented surface holonomy. The first idea that comes to one's mind is
to replace 2-forms on $M$ by 2-densities on $M$. These are 2-forms on the orientation cover
$\hat M$ of $M$ that are odd under the orientation reversion involution $\sigma$ of $\hat M$,
i.e.\ $\sigma^*\omega=-\omega$. Remembering that 2-forms are in fact objects in the
2-category in the third entry of Example \ref{ex1}, we realize that we should better not  impose equality
of objects on the nose, but only up to isomorphism. This is also clear from a more physical
point of view: we should impose equality only up to gauge transformations.

To set up the correct category $JGrbtriv^\nabla(M)$ of trivial objects for unoriented surface
holonomy, we first introduce the category $JBun^\nabla(M)$ of Jandl bundles: its objects are pairs,
consisting of a $\mathrm{U}(1)$-bundle $P\to M$ and a smooth map $\sigma:M\to
{\mathbb Z}_2$. Morphisms $(P,\sigma)\to (Q,\mu)$ of Jandl bundles only exist if $\sigma=\mu$.
The definition
$$ (P,\sigma)\otimes(Q,\mu):= (P\otimes Q^\sigma,\sigma\mu)$$
endows this category with the structure of a monoidal category. Here $Q^\sigma$
is the $\mathrm{U}(1)$-bundle obtained by taking the inverse of the $\mathrm{U}(1)$-bundle on the connected components
on which $\sigma$ takes the value $-1\in{\mathbb Z}_2$.

We are now ready to define the bicategory $JGrbtriv^\nabla(M)$ of trivial Jandl gerbes:
\begin{itemize}
\item
Its objects $I_\omega$ are in bijection to 2-forms $\omega\in\Omega^2(M)$.
\item 
Its 1-morphisms $I_\omega\to I_{\omega'}$ are Jandl bundles $(P,\sigma)$
with curvature $\mathrm{Curv}(P)=\sigma \omega'-\omega$.
\item
Its 2-morphisms are morphisms of Jandl gerbes which have to obey suitable coherence
conditions.
\end{itemize}

Trivial Jandl gerbes form a 2-prestack on the category of smooth manifolds. We can apply the 
general stackification procedure of Theorem 3.3 of \cite{tncs} and obtain a 2-stack of Jandl gerbes.
One can describe a Jandl gerbe on a smooth manifold $M$  explicitly in terms of descent data:
a surjective submersion $Y\to M$, a 2-form $\omega\in \Omega^2(Y)$, a Jandl bundle
$(P,\sigma)$ on $Y\times_MY$ and a certain morphism of Jandl bundles. Since morphisms
of Jandl bundles only exist in the case of equality of the $\mathbb{Z}_2$-valued functions
$\sigma$, the data $(Y,\sigma)$ combine into local data describing a $\mathbb{Z}_2$-cover 
$\mathcal{O}(\mathcal{G})$
on $M$, the orientation cover of the Jandl gerbe. In string theory, the orientation cover is 
sometimes called
the target space; the natural involution on the orientation cover $\mathcal{O}(\mathcal{G})$
is then called the orientifold involution.

One can also derive an explicit formula for the Wess-Zumino term on an unoriented surface
for a given Jandl gerbe. It can be explicitly written as a combination of integrals
of 2-forms and 1-forms and the evaluation of $\mathrm{U}(1)$-valued functions. 
Rewriting this formula on the orientation cover yields a formula obtained earlier in
\cite{ssw}. Checking that
this formula does not depend on the choices involved is a nice exercise using Stokes' theorem.
\footnote{
We are sure that Max would have liked this exercise much more than the framework of bicategories
that has lead us to the formula.} 
One can also compare Jandl gerbes on compact Lie groups
to algebraic results and finds complete agreement.

\section{RCFT correlators from TFT}\label{tncs_sec4}

\subsection{Surface operators}

In this last section, we first briefly move to 3-categories, keeping a rather informal
style of exposition. In fact, many new
developments (see e.g.\ \cite{bdh}) point to the direction that Chern-Simons theories or
chiral conformal field theories with fixed central
charge naturally form a 3-category. Similar comments apply to modular tensor categories.
Since we want to avoid the use of 3-categories, we will  later on restrict to endomorphisms of a fixed
object in the 3-category which have the structure of a bicategory.

In the case of three-dimensional topological field theories, this 
higher categorical structure nicely fits with the structure of 
defects of various codimension: consider a three-manifold $M$ that is separated by
an embedded surface $\Sigma\subset M$ into two disjoint parts, $M=M_1\sqcup \Sigma\sqcup M_2$.
Suppose that a Chern-Simons theory of type ${\mathcal C}_i$ is defined on each open submanifold $M_i$.
In this situation, transition conditions at the interface $\Sigma$ have to be specified. Such a set
of conditions  has been called a surface operator in \cite{kasa}. 

Let us make surface operators slightly more concrete for the case when the topological field
theories on $M_i$ are abelian Chern-Simons theories and thus
given by even lattices $L_i\subset {\mathbb R}^n$. Then (types of) topological surface
operators are in bijection to Lagrangian subgroups $S\subset L_1^* /L_1\times \overline{ L_2^*/ L_2}$,
where $L^*$ denotes the lattice dual to the lattice $L$.

Two surface operators $S_1$ and $S_2$ between the same pair of Chern-Simons theories
${\mathcal C}_1$ and ${\mathcal C}_2$ might in turn be separated by a defect of
codimension 2, a line. This
is in fact a generalization of the well-known concept of a Wilson line: in case the
two Chern-Simons theories are identical, ${\mathcal C}_1={\mathcal C}_2$, there is
a distinguished surface operator, the invisible surface operator $\mathrm{Inv}$.
It corresponds to the identity 1-morphism of ${\mathcal C}_1$ in the underlying 
3-category.
In the case of an abelian Chern-Simons theory, it corresponds to the diagonal subgroup
$S\subset L \times \overline L$. Line operators in the invisible surface operator are just
ordinary Wilson lines. 

We thus arrive at a (monoidal) bicategory that can be assigned to any Chern-Simons theory or, more
generally, to any chiral conformal field theory or modular tensor category: its objects 
are topological surface operators, 1-morphisms are generalized Wilson lines, 2-morphisms 
label insertions on the generalized Wilson lines.

\subsection{The Kreuzer-Schellekens bihomomorphism}

We make this picture more concrete in the following situation: consider a generalized
Wilson line $W$ separating the invisible surface operator for a modular tensor category
$\mathcal C$ from some general surface operator $S$. One can argue \cite{kasa2} that
this corresponds to a Frobenius algebra $A_W$ in the modular tensor category $\mathcal C$.
A Frobenius algebra is an object $A$ in $\mathcal C$, together with a unital associative
multiplication $\mu: A\otimes A\to A$ and a counital coassociative comultiplication
$\Delta: A\to A\otimes A$ such that $\Delta$ is a morphism of $A$-bimodules.
For pictures, we refer to section 2.2 of \cite{kasa2}

For a general tensor category, even for a modular tensor category, Frobenius algebras
are hard to classify.
There is however a systematic construction of a subclass of such algebras \cite{paper3}. 
Choose a subgroup $H$ of the Picard group $\mathrm{Pic}(\cal{C})$ 
of the modular tensor category $\mathcal C$. This is the finite abelian
group consisting of isomorphism classes of invertible objects in $\mathcal C$, in other
words of simple currents \cite{schellekens}.
The associativity constraint $\mathcal C$ on the category yields a closed 3-cochain 
$\psi$ on the Picard group $\mathrm{Pic}(\cal{C})$ and, by restriction, on the subgroup $H$. Any
2-cochain 
$$\omega: H\times H \to {\mathbb C}^\times $$
such that ${\mathrm d}\omega=\psi$ gives the structure of an associative
algebra on the object $\oplus_ {h\in H} U_h$ of $\mathcal C$. Two
cochains $\omega$ and $\omega'$ that differ by an exact cochain,
$\omega'-\omega={\mathrm  d}\eta$ for some 1-cochain $\eta$ give rise
to isomorphic algebras.

Let us first assume that the associator $\psi$ is trivial. In this case,
isomorphism classes of such algebras are classified by
the cohomology group $H^2(A,{\mathbb C}^\times)$ for
the abelian group $A=H$. It is a classical result that this
group is isomorphic to the group $AB(A,{\mathbb C}^\times)$ of alternating complex-valued
bihomomorphisms on the group $A$ via
$$\begin{array}{rll}
H^2(A,{\mathbb C}^\times) &\stackrel\sim\to & AB(A,{\mathbb C}^\times) \\
{}[\omega]&\mapsto & \xi(g,h)= \frac{\omega(g,h)}{\omega(h,g)} 
\end{array} $$

For the Picard subcategory of a modular tensor category, the associator $\psi$ is,
in general, non-trivial. Max Kreuzer and Bert Schellekens realized \cite{krsc}
in which way the condition on the bihomomorphism to be alternating has to be
modified; the corresponding bihomomorphisms have therefore been called Kreuzer-Schellekens
bihomomorphisms \cite{paper3}.

An alternating bihomomorphism $\xi\in AB(A,{\mathbb C}^\times)$ gives rise to
a twisted group group algebra ${\mathbb C}_\xi A$, i.e.\ a complex algebra 
with distinguished basis $(b_a)_{a\in A}$ in which the group law of $A$ is
realized only projectively, $b_{a_1}\cdot b_{a_2}= \omega(a_1,a_2) b_{a_1\cdot a_2}$.
The way a Kreuzer-Schellekens bihomomorphism on a subgroup $A$ of the Picard group of
a tensor category $\mathcal C$ gives rise to a (special symmetric) Frobenius algebra
$A_\xi$ in $\mathcal C$ is a generalization of this classical situation to a braided
setting.

Kreuzer-Schellekens algebras $A_\xi$ have the important property that on the
corresponding surface operator, one can punch out holes for free. Pictorially,

\begin{figure}[h!]
\centerline{\scalebox{1.0}{\includegraphics{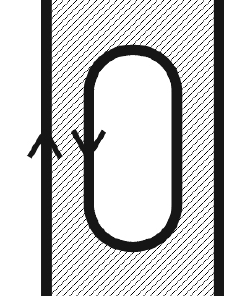}}
$=$
\scalebox{1.0}{\includegraphics{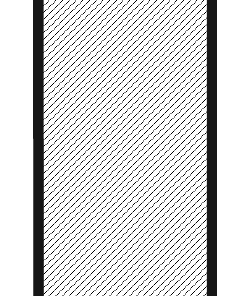}}
}
\caption{Punching out a hole}
\label{punch_fig}
\end{figure}

This follows algebraically from the fact that the coproduct $\Delta$ of a Kreuzer-Schellekens
algebra $A_\xi$ can be normalized ins such a way that it is a right inverse to the product,
$\mu\circ\Delta = \mathrm{id}_{A_\xi}$. In particular, Kreuzer-Schellekens algebras are
separable algebras in $\mathcal C$.

\subsection{RCFT correlators and the Kreuzer-Schellekens partition function}

Another important result in rational conformal field theory due to Max Kreuzer and
Bert Schellekens \cite{krsc} is the most general form \cite{krsc} of a modular 
invariant partition function of simple current type.

We place this result into the context of more recent developments. The partition function
is just one  specific correlator; by now the TFT approach to RCFT correlators provides 
a model-independent construction of all correlators (see e.g.\ \cite{icm} for a review). 
Let $\Sigma$ be an
oriented world sheet. Suppose that the chiral data for the rational conformal field theory
in question are described by the modular tensor category $\mathcal C$. According to
the results of Reshetikhin and Turaev, there is a three-dimensional topological field theory 
$$ \mathrm{tft}_{\mathcal C} : \mathrm{Cobord}_{3,2,1}\to 2\mathrm{Vect} $$
based on the modular tensor category.
The principle of holomorphic factorization states that
the correlator $\mathrm{Cor}(\Sigma)$ is an element in the space of conformal
blocks $\mathrm{tft}_{\mathcal C}(\hat\Sigma)$ which is assigned by the topological field theory 
to the orientation double $\hat \Sigma$ of $\Sigma$. The vector $\mathrm{Cor}(\Sigma)$ 
has to be invariant under the action of the mapping class group $\mathrm{Map}(\Sigma)$ and to obey
factorization constraints.

Let us sketch the TFT construction of RCFT correlators; for simplicity, we restrict to the case
when the oriented worldsheet $\Sigma$ has empty boundary, $\partial\Sigma=\emptyset$. In this
case, we consider the three-manifold $M_\Sigma:= \Sigma\times [-1,1]$ with a surface
operator placed at the surface $\Sigma\times\{0\}$. Generalized Wilson lines are
placed along the intervals $\{p\}\times [-1,1]$ for each bulk insertion at a point
$p\in\Sigma$. Since $M_\Sigma$ can be seen
as a cobordism $\emptyset\to \Sigma \sqcup\overline\Sigma\cong\hat\Sigma$, applying the
tft functor gives an element
$$ Cor(\Sigma):=\mathrm{tft}_{\mathcal C}(M_\Sigma) 1 \in 
\mathrm{tft}_{\mathcal C}(\partial M_\Sigma) =
\mathrm{tft}_{\mathcal C}(\hat\Sigma) \,\,  $$
that depends, of course, on the choice of the surface operator on the surface $\Sigma\times\{0\}$
in $M_\Sigma$.
Suppose that the surface operator has the property that one can punch out 
holes for free, as in figure 2. Then, by punching enough holes, we reduce the surface
operator to a network of strips and replace it effectively  by the dual of a 
triangulation of $\Sigma\times\{0\}$. This triangulation is
labeled with the Frobenius algebra $A_W$ corresponding the generalized Wilson line $W$
separating the surface operator and the invisible surface operator $\mathrm{Inv}$.
This yields an expression of the RCFT correlator in terms of the invariant of a
three-manifold $M_\Sigma$ with an embedded ribbon graph to which the Reshetikhin-Turaev
construction can be applied.

It can now be shown \cite{fjel1} that the elements $Cor(\Sigma)$ are invariant under the 
action of the mapping class group of $\Sigma$ and obey factorization constraints. 
I can even be shown that any consistent set of correlators for a given chiral rational conformal
field theory is of this form \cite{fjel2}.

The formalism is also well suited to explicit calculations: one can compute
structure constants and partition functions for boundary fields, bulk fields and
defect fields. One can apply the RCFT construction in particular to the case of a
Kreuzer-Schellekens algebra and compute the partition function of bulk fields.
One obtains exactly the result first derived by Max Kreuzer and Bert Schellekens,
based in particular on Max Kreuzer's careful understanding of orbifold theories
(from which also one of the present authors benefitted a lot \cite{poinc}).

\bigskip

We hope that this contribution helps to convince practitioners of
field theory that bicategories - and higher categories - are one helpful tool
to solve problems in
topological, classical and quantum field theories. Max Kreuzer was a fine
practitioner of field theory with a keen appreciation of the true problems arising
in these theories. He might have looked at bicategories with a somewhat skeptical smile.
We like the idea that he would have tried immediately to translate some of the insights 
sketched in this contribution in a language closer to his heart - 
and we gratefully remember that this was frequently his way to show his appreciation for
the results of his colleagues.

\section*{Acknowledgments}
TN and CS are partially supported by the Collaborative Research Centre 676 ``Particles,
Strings and the Early Universe - the Structure of Matter and Space-Time'' and the
cluster of excellence ``Connecting Particles with the Cosmos''.  CS is partially supported
by the Research Priority Program SPP 1388 ``Representation theory''.

\bibliographystyle{ws-rv-har}

%\bibliography{ws-rv-sample}

\end{document}